\documentclass[lettersize,journal]{IEEEtran}
\usepackage{amsmath,amssymb,amsfonts}
\usepackage{algorithmic}
\usepackage{algorithm}
\usepackage{array}
\usepackage[caption=false,font=normalsize,labelfont=sf,textfont=sf]{subfig}
\usepackage{textcomp}
\usepackage{stfloats}
\usepackage{url}
\usepackage{verbatim}
\usepackage{graphicx}
\usepackage{cite}
\usepackage{graphicx}
\usepackage{booktabs}
\usepackage{multirow}
\usepackage{xcolor}
\usepackage{siunitx}
\usepackage{lipsum}
\usepackage{ifthen}
\usepackage[normalem]{ulem}
\definecolor{darkgreen}{RGB}{0,100,0}

\DeclareSIUnit{\rad}{rad}
  
\begin{document}

\title{LIVE-RIS: Real-Time In-Flight Actuation of UAV-Mounted RIS} 

\author{David Müller, Kevin Weinberger, Aydin Sezgin,~\IEEEmembership{Senior Member,~IEEE,} Martin Mönnigmann,~\IEEEmembership{Member,~IEEE,}
\thanks{This work was funded by the German Research Foundation (DFG) in the course of the project SPP2433 under the project no. 541021107 (Measurement Technology on Flying Platforms) under grant MO 1086/17-1, grant SE 1697/22-1, and grant INST 213/1028-1 FUGB.

David Müller and Martin Mönnigmann are with the chair of Automatic Control and Systems Theory, Department of Mechanical Engineering, Ruhr-Universität Bochum, 44801, Bochum, Germany {\tt\small david.mueller-r21, martin.moennigmann@ruhr-uni-bochum.de}.

Kevin Weinberger and Aydin Sezgin are with the chair of Digital Communication Systems, Department of Electrical Engineering, Ruhr-Universität Bochum, 44801, Bochum, Germany {\tt\small kevin.weinberger, aydin.sezgin@ruhr-uni-bochum.de}.}
\thanks{
}}

\markboth{Submitted to IEEE TRANSACTIONS ON INTELLIGENT VEHICLES}%
{Shell \MakeLowercase{\textit{et al.}}: A Sample Article Using IEEEtran.cls for IEEE Journals}

\IEEEpubid{
}

\maketitle

\begin{abstract}

Reconfigurable intelligent surfaces (RIS) are emerging as a key technology for sixth-generation (6G) wireless networks due to their ability to dynamically control the propagation environment.
To ensure favorable Line-of-Sight (LoS) conditions in real-world applications, the RIS is mounted on an unmanned aerial vehicle (UAV).
While the potential of UAV-mounted RIS has been extensively studied in theoretical works, experimental validation with real-world data remains limited.
Such validation is particularly important, as UAV motion and disturbances may degrade the performance of the RIS-enabled link.
In this paper, we present the first fully functional, real-time capable UAV-mounted RIS prototype and validate its performance through experimental measurements under realistic disturbances and hardware constraints.
We show that the RIS pose can be predicted based on the UAV's extended Kalman filter (EKF) and onboard sensors.
By utilizing this estimation, we demonstrate that the RIS can be reconfigured in real time, effectively mitigating disturbance effects and preserving the performance gains of the RIS-enabled link.
Furthermore, we systematically evaluate different deployment locations to provide insights into RIS performance in real-world scenarios.
\end{abstract}


\section{Introduction}\label{sec:Introduction}
\IEEEPARstart{R}{econfigurable} Intelligent Surfaces (RISs) are emerging as a key enabler of 6G wireless networks, transforming the radio environment into a programmable entity and facilitating core 6G visions such as ubiquitous connectivity, ultra-high energy efficiency, and integrated sensing and communication \cite{6G_modernVision}. By adjusting the phase, amplitude, and frequency of incoming signals at each reflecting element, RISs can enhance network performance, improve resilience~\cite{Weinberger2023}, and even facilitate new applications~\cite{RISApp}. Therefore, proper RIS configurations are essential to meet specific use case requirements.
In addition to the chosen configuration, the RIS’s deployment position also plays a major role, due to the multiplicative path loss caused by the transmitter (Tx)-RIS-receiver (Rx) link. 
Due to the mobility of users in wireless networks,  these requirements are not always met, especially when considering blockages and changing terrain heights. To circumvent this problem, unmanned aerial vehicles (UAVs) have been proposed to carry the RIS, so that a line-of-sight (LoS)-link can be guaranteed~\cite{Weinberger2023abc}. 
Because of its low weight and power efficiency, RISs are excellent alternatives to heavy and power-hungry base stations for use with UAVs.
While mounting a RIS on a UAV enables it to reposition itself, 
the UAV is subject to external disturbances and internal sources of uncertainty.
These uncertainties lead to constant changes in the position~\cite{Mueller2025} and orientation~\cite{Mueller2024} of the UAV, even when hovering at a fixed position.
Consequently, the RIS-reflected path is subject to these changes, affecting signal transmission.

The benefits of UAV-mounted RIS have been widely investigated in the literature. Their application extends beyond establishing LoS-links, as they can significantly expand wireless coverage, enhance communication performance and security, and support efficient data collection~\cite{Pogaku2022}.
However, the majority of existing works rely exclusively on numerical simulations to validate their findings, mostly neglecting the impact of real-world constraints and disturbances on the channel link.
Such factors can be relevant in practice, as they may impact channel quality and potentially limit the practical applicability of purely theoretical results~\cite{Weinberger2024b}.

Consequently, the existing literature still lacks real-world insights and experimental data, since no active UAV-mounted RIS has yet been deployed and studied in practice. Motivated by this fundamental gap, we present, to the best of our knowledge, the first UAV-mounted RIS that performs real-time reconfiguration in response to environmental changes and validate it through extensive real-world experiments. 
Our work provides detailed insights into the developed prototype, highlighting hardware constraints that have largely been overlooked in prior studies.
Furthermore, we evaluate the performance of the UAV-mounted RIS across multiple scenarios and deployment locations, thereby assessing its practical utility under dynamic, real-world conditions.

\section{Related Work}
\label{sec:RelatedWork}
The concept of UAV-mounted RIS has been investigated for a wide range of applications, including wireless coverage extension, communication performance enhancement, physical-layer security provisioning, and wireless data collection~\cite{Pogaku2022 ,Deng2025, Khan2022}.
Stationary deployment scenarios, where the UAV hovers at a fixed position and the transmitter and receiver are static or exhibit low mobility have been extensively studied in the literature~\cite{Chen2026,Tyrovolas2023,Wei2023,Tang2021,Mahmoud2021,Guo2023}.
For example,~\cite{Tyrovolas2023} analyzes the coverage probability of randomly deployed sensors with a hovering UAV for efficient data collection, while~\cite{Wei2023} proposes an aerial RIS-assisted secure transmission design in wireless networks, where the UAV-mounted RIS positioning is optimized to enhance the propagation environment.

Recent research has explored dynamic deployment scenarios in which the UAV continuously repositions the RIS or follows complex trajectories~\cite{Eskandari2023, You2021, Tang2023, Hao2025}.
Such approaches include optimal trajectory planning and path learning utilizing neural networks, both based on the characteristics of the communication link.
In~\cite{Eskandari2023}, the authors propose a two-stage model predictive control (MPC)-based 3D navigation framework for UAV-mounted RIS for establishing an uninterruptable LoS link.
In the first stage, feasible navigation boundaries are determined based on the UAV’s energy consumption while accounting for UAV limitations and obstacles.
In the second stage, the LoS service is formulated as an optimization problem to generate a UAV trajectory subject to the constraints obtained in the first stage.
The authors in~\cite{Tang2023} formulate an optimization problem to maximize the average secrecy rate by jointly optimizing the RIS reflection coefficients and the UAV trajectory.
The problem is decomposed into two layers, with the trajectory planning solved via reinforcement learning.

The concept of UAV-mounted RIS swarms has also been explored to further enhance network performance through coordinated RIS deployment~\cite{Shang2023,Liao2022,Nguyen2022,Deng2024}.
For instance,~\cite{Nguyen2022} proposes a novel UAV-RIS-assisted communication scheme for network coverage extension in a massive MIMO system.
The total network throughput is maximized through joint optimization of the power allocation coefficients at a massive MIMO base station and the phase shifts of multiple UAV-mounted RIS.

However, all of these works validate their results solely through numerical simulations, mostly neglecting real-world constraints such as internal (e.g., sensor noise) and external disturbances (e.g., wind) and UAV position or orientation uncertainties.
In most cases, even practical RIS hardware limitations, such as discrete phase shifts, are not considered.

A few studies~\cite{Mueller2024,Weinberger2025,Mueller2025} overcome this limitation by including actual UAV flight data in their numerical simulations, enabling a closer approximation of real-world performance.
In~\cite{Mueller2024}, the authors investigate UAV orientation uncertainties caused by internal disturbances, e.g. EKF linearization and sensor measurement errors, and analyze their impact on the channels of UAV-mounted RIS.
Subsequently,~\cite{Weinberger2025} propagates these uncertainties through the UAV-mounted RIS channel model to gain additional insights of their influence on channel behavior.
Similarly,~\cite{Mueller2025} investigates UAV position uncertainties while accounting for external disturbances such as wind.
Although these studies demonstrate that a continuously reconfigured UAV-mounted RIS can significantly outperform a passive RIS, even in the presence of position or orientation uncertainties and under some practical RIS limitations (e.g., discrete phase shifts), their results remain simulation-based.
They largely neglect uncertainties arising from system integration, such as delays, hardware impairments, and real-world effects on signal propagation.

To the best of our knowledge, only one work has experimentally demonstrated the feasibility of UAV-mounted RIS.
In~\cite{Weinberger2023abc}, the authors predetermined the optimal RIS configuration for a reference point and maintained it during flight with a real-world UAV-mounted RIS prototype.
Their results show that reaching the reference point with an optimal RIS configuration significantly enhances signal transmission compared to a passive (turned-off) RIS.
However, optimizing the RIS for a single reference point does not exploit the full potential of the proposed system.
In particular, the results from~\cite{Mueller2024,Weinberger2025,Mueller2025} show, that an optimal, yet static RIS configuration is insufficient in the presence of disturbances and uncertainties.

In summary, existing works on UAV-mounted RIS primarily rely on theoretical analyses and simulations, often under idealized assumptions, while practical implementations and experimental validations remain largely unexplored.
To bridge this gap, this work presents the first real-time capable UAV-mounted RIS and provides an experimental evaluation under realistic conditions, thereby advancing the understanding of RIS performance beyond purely theoretical studies.

\section{System Model}
\label{sec:SystemModel}
The system model in this paper assumes a single-antenna Tx transmitting signals towards a single-antenna Rx. It is further assumed that the direct Tx-Rx link is blocked and a RIS-equipped UAV is used to establish a communication link. The deployed RIS consists of $M$ reflecting elements, each capable of altering the phase of the reflected signal. The reflected signals are then received by the Rx, resulting in an effective channel between Tx-RIS-Rx as
\begin{align}\label{eq:heff}
    h^\mathsf{eff} = \sum_{m=1}^{M} h_m \theta_m g_m, \\[-21pt] \nonumber
\end{align}
where $h_m$ and $g_m$ represent the channel coefficients between
$m$-th RIS element and the Tx/Rx, respectively. Further, the reflect coefficient $\theta_m = A_m(\varphi_m)e^{j\varphi_m}$ is defined by the tunable phase  $\varphi_m\in[0,2\pi)$ induced to the reflected signal at reflecting element $m$ as well as the phase-dependent reflect amplitude $A_m \in [0,1]$.
The cascaded channel coefficient between Tx-RIS-Rx over the \( m \)-th element can consequently be formulated as $h^\mathsf{casc}_m = h_m g_m$.

We determine the cascaded channel components for each reflect element \( m \) based on the geometry of the LoS path, where \( d^h_m \) (\( d^g_m \)) is defined as the distance between the Tx (Rx) and the \( m \)-th RIS element. The cascaded Tx-RIS-Rx channel through the \( m \)-th RIS element is given by~\cite{goldsmith2005wireless}
\begin{align}\label{eq:chanModel}
     h^\mathsf{casc}_m = h_m g_m = \left[ \frac{c}{4\pi \nu d^h_m} e^{j\frac{2\pi}{\lambda}d^h_m} \right] \left[ \frac{c}{4\pi \nu d^g_m} e^{j\frac{2\pi}{\lambda}d^g_m} \right],
\end{align}
where \( c \) is the speed of light, \( \nu  \) the carrier frequency and \( \lambda \) the corresponding wavelength.

\section{System Integration}
\label{sec:SystemIntegration}
The UAV-mounted RIS system used in this paper, can be decomposed into two primary subsystems: the RIS subsystem and the UAV subsystem. Each subsystem is described in detail in the following subsections. The communication interface between both subsystems is addressed separately.

\subsection{RIS Subsystem}
\label{sec:RISsetup}
\begin{figure}[t]
	\centering
	\includegraphics[width=1\linewidth] {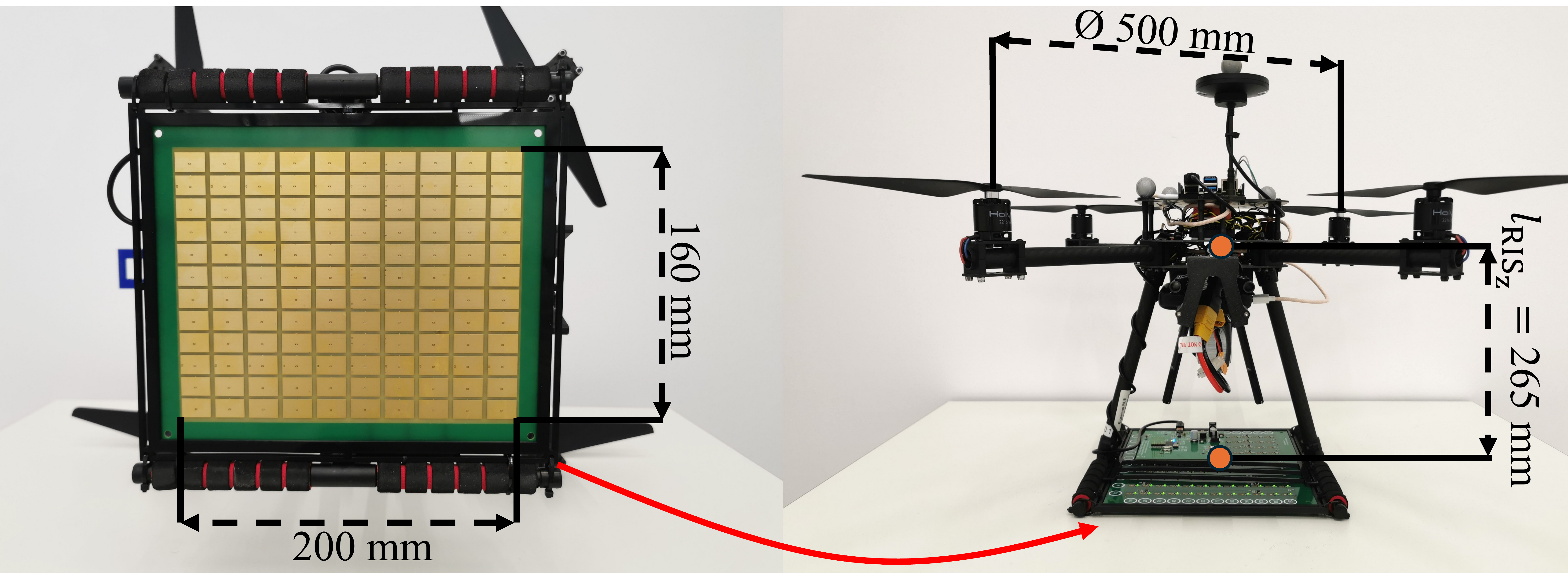}
	\caption{Left: RIS prototype and its dimensions. Right: RIS prototype mounted to a customized Holybro X500. Orange dots are representing the origin of the UAV’s body frame (top) and the center of the RIS (bottom).}
	\label{fig:UAVSetup}
\end{figure}

The RIS prototype\footnote{https://github.com/mheinri/OpenSourceRIS} has dimensions of 20$\times$\SI{16}{\centi\meter} and consists of $M = 120$ elements arranged in a 10$\times$12 array (see Fig.~\ref{fig:UAVSetup}).
Each element is equipped with a PIN diode to enable 1-bit discrete phase shifting.
The RIS is designed as a binary-switching surface with a \SI{180}{\degree} phase shift at the designed carrier frequency of $\nu=5.385$\si{\giga\hertz}.
During phase shifting, the reflected signal experiences a \SI{3}{\deci\bel} attenuation due to hardware impairments in the phase-shifting circuitry.
We use a Raspberry Pi 4B as the RIS controller, which is connected to the RIS via a serial port with a baud rate of 115200 Bd.
The maximum reconfiguration rate of the RIS, i.e., the frequency at which the elements’ phase shifts can be updated, is \SI{50}{\hertz}.

\subsection{UAV Subsystem}
\label{sec:UAVSetup}

The UAV carrying the RIS is a customized Holybro X500 Quadcopter with a rotor-to-rotor diameter of 500\si{\milli\meter} (see Fig.~\ref{fig:UAVSetup}), which can carry payload of up to 1\si{\kilogram}.
The UAV is controlled by a Pixhawk Cube Orange flight controller.
The flight controller holds three different inertial measurement units (IMU), where three IMUs are used for redundancy (the primary IMU is a ICM20602, the secondary one a ICM20948 and the third one a ICM20649\footnote{https://ardupilot.org/copter/docs/common-thecubeorange-overview.html}).
The open-source software ArduPilot\footnote{https://ardupilot.org/} (V4.6.3) is installed on the flight controller, running an individual EKF for every IMU. This enables the UAV to use lane-switching, a transition between multiple EKFs in case variances exceed a
threshold\footnote{https://ardupilot.org/dev/docs/common-ek3-affinity-lane-switching.html}. The active EKF is selected in the same order as the IMUs. The primary and secondary IMUs are mounted on a separate temperature-controlled, vibration-isolated board, while the third IMU is located directly on the flight controller.
Three different IMUs are installed on the flight controller, as they have different frequency responses and will therefore respond differently to vibrations\footnote{https://docs.cubepilot.org/user-guides/autopilot/the-cube-module-overview}.
We tuned filter parameters for better filtering of vibrations by running an in-flight Fast Fourier Transform on the UAV and analyzing spectra for characteristic frequencies.
Measurements of the IMU are then fused with other sensors by the EKF.

Since the experiments are conducted inside a laboratory, GNSS position information is not available. We replace the position measurements received by the GNSS antenna with position measurements from a Motion Capture System (MCS) to enable autonomous indoor flight.
We use an ESP8266 microcontroller connected to the flight controller running MAVESP8266\footnote{https://github.com/BeyondRobotix/mavesp8266} firmware to send the position information via an additional WiFi connection to the UAV during flight.
The EKF processes the MCS measurements with a frequency of \SI{5}{\hertz}, equivalent to the nominal GNSS update rate.
Details on the motion capture system are provided in Sec.~\ref{sec:Experiments}.
Similar to real-time kinematic (RTK) GNSS setups, we also substitute the on-board barometer and magnetometer measurements with MCS measurements. Indoor laboratory environments typically degrade barometric altitude measurements and introduce substantial
magnetic disturbances caused by reinforced concrete structures, compromising reliable state estimation.

The RIS is mounted under the UAV, where its center is located $l_{\mathrm{RIS}_z} = 265$\si{\milli\meter} underneath the origin of the UAV's body frame (see Fig.~\ref{fig:UAVSetup}), with no offsets in the $x$-$y$-plane.
The Raspberry Pi is mounted on top of the UAV, with almost no offset in $z$-direction, and no offsets in the $x$-$y$-plane.

\subsection{Communication between Subsystems}
\label{sec:Communication}

The flight controller of the UAV communicates with the RIS controller through a serial interface with a baud rate of 500000 Bd.
Data from the flight controller to the RIS controller is streamed via the MAVLINK 2 protocol at a rate of 50\si{\hertz}, matching the sampling frequency of the RIS.
To reduce latency and ensure consistent data transmission, only the UAV’s estimated position, velocity, and orientation, together with the most recent IMU measurements (linear acceleration and angular velocity along each axis), are transmitted.
In Sec.~\ref{sec:RISPosePred}, we show that these information are sufficient for accurate RIS pose prediction.

\section{RIS Actuation}
\label{sec:RISActuation}

RIS actuation is fully handled by the RIS controller. By partitioning the system into two independent subsystems, the computational load on the flight controller is reduced, thereby enabling stable flight operation independent of the RIS subsystem.
The RIS actuation is divided into two stages. The first stage focuses on RIS pose prediction.
As shown in (\ref{eq:chanModel}), the RIS configuration depends on the accurate distances between Tx-RIS and RIS-Rx, making precise pose prediction essential.
To minimize power consumption and extend UAV flight time, no additional sensors are introduced. Instead, we rely solely on the UAV's EKF estimates and the previously introduced onboard sensors.
We employ a lightweight prediction approach to keep the computational load low and maintain short prediction horizons.
The second stage determines the optimal RIS configuration based on the predicted position and orientation. This requires formulating an optimization problem that enables real-time actuation while accounting for hardware constraints and impairments.

\subsection{RIS Pose Prediction}
\label{sec:RISPosePred}

Precise pose prediction of the RIS becomes essential, as latency arises both from computing and applying the optimal RIS configuration and from communication delays between system components.
Consequently, the most recent EKF estimates cannot be used directly, as they are already outdated at the time of RIS actuation.
To address this, we employ an open-loop forward propagation, where the latest EKF estimate serves as the initial state for propagation over a prediction horizon.
The time interval between the most recent EKF estimate of the UAV and the application of the optimized RIS configuration is denoted by $t_{\mathrm{pred}}$, which is system-dependent and assumed to be constant in this work.
Accordingly, the required pose estimate corresponds to the prediction time instant $t_{k+p} = t_k + t_{\mathrm{pred}}$, with $k$ denoting the index of the time step.

The UAV's EKF pose and velocity estimates as well as the IMU data are extracted from the flight controller by the RIS controller at a frequency of \SI{50}{\hertz}.
The EKF data consists of the UAV position $p^w_{k,\mathrm{UAV}}\in \mathbb{R}^{3}$, the UAV velocity $v^w_{k,\mathrm{UAV}}\in \mathbb{R}^{3}$, and the UAV attitude $\Phi_{k,\mathrm{UAV}},\Theta_{k,\mathrm{UAV}},\Psi_{k,\mathrm{UAV}}$ in the NED frame.
The IMU data comprises the acceleration $a^b_{k,\mathrm{IMU}}\in \mathbb{R}^{3}$ along each spatial axis, as well as the angular velocity $\omega^b_{k,\mathrm{IMU}}\in \mathbb{R}^{3}$ about each axis at the current time step $k$.
The superscripts $w$ and $b$ indicate whether the quantities are expressed in the world frame or in the UAV body frame, respectively.
The IMU data is preprocessed by the flight controller using a low-pass filter, thereby suppressing oscillations caused by airframe vibrations.

Prior to predicting the UAV pose, we estimate the IMU bias for both the gyroscope, $\omega^b_{\mathrm{bias}}$, and the accelerometer, $a^b_{\mathrm{bias}}$, during the initial stationary phase of the UAV to enable bias correction of the most recent IMU measurements.
Additionally, we allow bias estimates whenever the UAV becomes stationary again.
We then compute the bias compensated measurements as
\begin{align} \label{eq:corrIMU}
    \omega^b_{\mathrm{corr}}=\omega^b_{k,\mathrm{IMU}}-\omega^b_{\mathrm{bias}}\\
    a^b_{\mathrm{corr}} = a^b_{k,\mathrm{IMU}}-a^b_{\mathrm{bias}}
\end{align}
removing sensor offsets.
We predict the attitude of the UAV at $t_{k+p}$ by deriving the rotation vector
\begin{equation} \label{eq:deltaRot} 
    [\Delta \Phi,\Delta \Theta,\Delta \Psi] = \omega^b_{\mathrm{corr}} t_{\mathrm{pred}}
\end{equation}
and map it into a quaternion increment $\Delta q_{\mathrm{pred}}$.
After converting the most recent attitude estimated by the EKF $\Phi_{k,\mathrm{UAV}},\Theta_{k,\mathrm{UAV}},\Psi_{k,\mathrm{UAV}}$ into a unit quaternion $q_{k,\mathrm{UAV}}$, we obtain the predicted attitude with
\begin{equation} \label{eq:q_config} 
    q_{k+p\mathrm{,UAV}} = q_{k\mathrm{,UAV}}\otimes\Delta q_{\mathrm{pred}}
\end{equation}
and normalize it into a unit quaternion.
Subsequently, we extract the predicted Euler angles $\Phi_{k+p\mathrm{,UAV}},$ $\Theta_{k+p\mathrm{,UAV}},\Psi_{k+p\mathrm{,UAV}}$.
For accurate UAV position prediction, we need to consider rotation-dependent acceleration during the integration interval~\cite{Forster2015}. Therefore, we employ midpoint integration as a trade-off between accuracy and computational efficiency.
We calculate the rotation vector with
\begin{equation} \label{eq:q_pred} 
    [\Delta \Phi_{k+\frac{p}{2}},\Delta \Theta_{k+\frac{p}{2}},\Delta \Psi_{k+\frac{p}{2}}] = \frac{1}{2}\omega^b_{\mathrm{corr}} t_{\mathrm{pred}}
\end{equation}
and derive the unit quaternion increment $\Delta q_{k+\frac{p}{2}}$.
The corresponding quaternion is given by
\begin{equation} \label{eq:q_pred} 
    q_{k+\frac{p}{2}} = q_{k,\mathrm{UAV}}\otimes\Delta q_{k+\frac{p}{2}},
\end{equation}
which is converted into Euler angles $\Phi_{k+\frac{p}{2}}$,$\Theta_{k+\frac{p}{2}}$,$\Psi_{k+\frac{p}{2}}$.
We transform the IMU accelerometer measurements into the world frame with
\begin{equation} \label{eq:q_pred} 
    a^w_{k,\mathrm{IMU}} = R^w_b(\Phi_{k+\frac{p}{2}},\Theta_{k+\frac{p}{2}},\Psi_{k+\frac{p}{2}})(a^b_{k,\mathrm{IMU}}-\varrho),
\end{equation}
where $R^w_b(\Psi_{k+\frac{p}{2}},\Theta_{k+\frac{p}{2}},\Phi_{k+\frac{p}{2}})\in \mathbb{R}^{3\times3}$ is the product of the rotation matrices
\begin{equation}
\label{eq:rotmatrix}
    R^w_b(\Psi_{k+\frac{p}{2}},\Theta_{k+\frac{p}{2}},\Phi_{k+\frac{p}{2}}) = R_z(\Psi_{k+\frac{p}{2}})R_y(\Theta_{k+\frac{p}{2}})R_x(\Phi_{k+\frac{p}{2}})
\end{equation}
and $\varrho$ represents the gravity vector.
The UAV position can then be predicted using the constant-acceleration kinematic equation
\begin{equation} \label{eq:p_pred} 
    p^w_{k+p\mathrm{,UAV}} = p^w_{k,\mathrm{UAV}}+v^w_{k,\mathrm{UAV}} t_{\mathrm{pred}}+\frac{1}{2}a^w_{k,\mathrm{IMU}} t_{\mathrm{pred}}^2.
\end{equation}
Based on the predicted UAV attitude and position, we can subsequently determine the RIS pose.
Due to the rigid connection between the UAV and the RIS, their predicted attitudes are identical
\begin{align}
\label{eq:rotRis}
    [\Phi_{k+p,\mathrm{RIS}},\Theta_{k+p,\mathrm{RIS}},\Psi_{k+p,\mathrm{RIS}}]\quad\,\,\,\,\,\,\,\,\,\, \\
    =[\Phi_{k+p\mathrm{,UAV}},\Theta_{k+p\mathrm{,UAV}},\Psi_{k+p\mathrm{,UAV}}].\nonumber
\end{align}
The position of the RIS $p^w_{k+p,\mathrm{RIS}}\in \mathbb{R}^3$ can be obtained with
\begin{align} \label{eq:p_RIS}
p^w_{k+p,\mathrm{RIS}}
&= p^w_{k+p,\mathrm{UAV}}  \nonumber\\
&\quad + R(\Psi_{k+p,\mathrm{RIS}},
           \Theta_{k+p,\mathrm{RIS}},
           \Phi_{k+p,\mathrm{RIS}})
           l_{\mathrm{RIS}}.
\end{align}
where $l_{\mathrm{RIS}}=[0,0,-l_{\mathrm{RIS}_z}]^T$ describes the spatial offset between UAV center of mass and the RIS and $R(\Psi_{k+p,\mathrm{RIS}},\Theta_{k+p,\mathrm{RIS}},\Phi_{k+p,\mathrm{RIS}})\in \mathbb{R}^{3\times3}$ is the product of the rotation matrices
as defined in \eqref{eq:rotmatrix}.
These quantities are subsequently transformed from the NED to the NWU frame, as both the RIS and the MCS used for ground-truth data operate in the NWU frame.
The corresponding transformation equations are omitted here, as this step is implementation-specific and not strictly required, since the RIS configuration can alternatively be adapted accordingly.

\subsection{RIS Optimization}
\label{sec:optProblem}
For optimization of the RIS configuration, we first derive the distances $d^h_m$ and $d^g_m$, i.e., between each reflecting element $m$ for the Tx-RIS and RIS-Rx link, respectively. 
Consequently, the cascaded channel coefficients $h^{\mathrm{casc}}_m$, which is prerequisite for the optimization process, can be determined for each predicted attitude \eqref{eq:rotRis} and position \eqref{eq:p_RIS} with (\ref{eq:chanModel}).

Since this optimization needs to be carried out in real time, we have to keep the computational effort in mind when devising the optimization algorithm.
We recall that the deployed RIS prototype is limited to applying only $180^\circ$ phase shifts, i.e., $\varphi_m\in\{0^\circ,180^\circ\}$, and suffers from a \SI{3}{\decibel} loss in reflection amplitude $A_m$ for $\varphi_m= 180^\circ$ (see Sec.~\ref{sec:RISsetup}).
Due to the blocked path between Tx and Rx, the phase of the effective channel $h^{\mathsf{eff}}$ can be chosen arbitrarily during the optimization process. 
This can be expressed as
\begin{equation} \label{optForm} 
    \varphi_m^*(C) = C - \varphi'_m, 
\end{equation}
where $C \in \mathbb{R}$ denotes an arbitrary value for the desired phase of $h^{\mathsf{eff}}$ and $\varphi'_m = \frac{2\pi}{\lambda}(d_m^h + d_m^g)$.
We select $C$ from a finite set to ensure that the optimization is fast enough for use in real time. In the present paper, we use the values
\begin{equation}\label{eq:quantizedPhaseShifts}
C\in {\mathcal{C}} = \{0, \frac{\pi}{4}, \frac{\pi}{2}, \frac{3\pi}{4}, \pi, \frac{5\pi}{4}, \frac{3\pi}{2}, \frac{7\pi}{4}\}, 
\end{equation}
which results in a suitable trade-off to identify a good solution within the available time~\cite{RIS_proto}.

After optimizing the phase shift $\varphi^*_m$ for every patch $m= \{1, \dots, M\}$, we need to determine which of the patch states, $\varphi_m^*(C)= 0^\circ$ or $180^\circ$, is the optimal choice.  
We choose a fast rounding operation in light of the real-time requirements.
For the resulting $|\mathcal{C}|$ candidate configurations, where $|\mathcal{C}|$ denotes the set's cardinality, we choose the best performing one by assessing and comparing the determined channel quality of the resulting RIS-facilitated links. 
This procedure can mathematically formulated as maximizing $|h_C^{\mathsf{eff}}|$ for all $C\in\mathcal{C}$ 
\begin{subequations}\label{eq:optProblem}
\begin{align}
\max_{C\in\mathcal{C}} |h_C^{\mathsf{eff}}| = & \Big|\sum_{m=1}^{M} {h}^{\mathsf{casc}}_m A_m(\text{rd}(\varphi_m^*(C))) \text{e}^{j \text{rd}(\varphi_m^*(C))}\Big| \label{optEq}\\[-6pt]
 \text{with}& \,\, A_m(\tau) = \begin{cases} $0.5012 (-3\text{dB})$ , \,\text{if } \tau = \pi \\ 1 , \quad\quad\quad\quad\,\,\, \,\,\,\,\:\text{otherwise}\end{cases}, \\[-6pt]
 &\quad \text{rd}(\tau) = \begin{cases} \pi , \quad\text{if } \frac{\pi}{2} \leq\tau < \frac{3\pi}{2} \\ 0, \quad\text{otherwise}\end{cases}\quad\,\,\,\,\,\,\,  , \label{rdFun} \\[-3pt]
 &\,\,\, C\in {\mathcal{C}} = \{0, \frac{\pi}{4}, \frac{\pi}{2}, \frac{3\pi}{4}, \pi, \frac{5\pi}{4}, \frac{3\pi}{2}, \frac{7\pi}{4}\}  ,\label{Theta} \\[-14pt] \nonumber
\end{align}
\end{subequations}
for every predicted pose at $t_{k+p}$. This optimization can be carried out efficiently, because the optimal $C$ and $\varphi^*_m(C)$ can be found independently for each patch by comparing the ${|\mathcal{C}|}= 8$ choices and rounding subsequently.

\section{Experimental Validation}
\label{sec:ExpVal}

In this section, we describe how the experiments are conducted. As the evaluation is based on real-world experiments, the data processing and synchronization are also briefly explained.
Subsequently, we evaluate the performance of the proposed prediction strategy, followed by an analysis of the performance gains achieved by actuating the UAV-mounted RIS in real-time.

\subsection{Measurement Scenarios and Experimental Setup}
\label{sec:Experiments}

All experiments are conducted in a flight lab under the same environmental conditions.
To demonstrate the feasibility of the proposed prototype, we consider scenarios with varying Tx, Rx, and UAV-mounted RIS positions.
We are treating hover flight in all scenarios, i.e. the UAV is commanded to maintain a fixed target position.
Trajectories toward the target position are excluded from evaluation.
In each scenario, the Tx and Rx antennas are oriented toward the target position of the UAV-mounted RIS.
As ground truth data is recorded in the NWU frame, all pose information are now given in the NWU frame.

\begin{figure}[ht]
	\centering
	\includegraphics[width=1\linewidth] {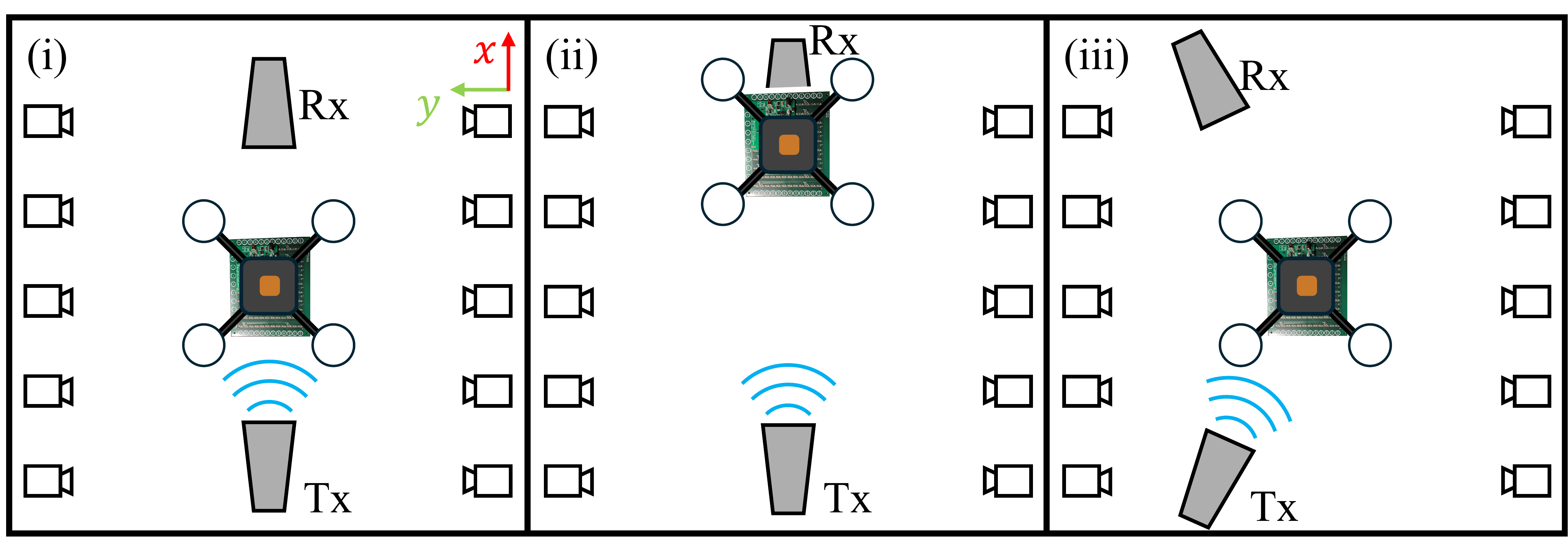}
	\caption{Schematic representation of the three scenarios considered for UAV-mounted RIS performance evaluation. From left to right: (i) RIS positioned at the midpoint between Tx and Rx, (ii) RIS positioned above the Rx, and (iii) RIS positioned off the direct path between Tx and Rx. In all scenarios, the RIS is deployed at a height of 2m.}
	\label{fig:scenarios}
\end{figure}

In the first scenario (i), the UAV-mounted RIS is positioned at the midpoint of the direct link between Tx and Rx.
The target position of the UAV-mounted RIS is set to $(0,0,2)$. The Tx and Rx are located at $(-1.36,0,0.7)$ and $(1.34,0,0.7)$, respectively.
In the second scenario (ii), the UAV-mounted RIS is positioned directly above the Rx, with the target position set to $(1.2,0,2)$. The Tx and Rx are located at $(-1.36,0,0.66)$ and $(1.55,0,0.75)$, respectively.
The third scenario (iii) treats a more unusual setup in which the UAV-mounted RIS is not located along the direct path between Tx and Rx.
Similar to scenario (i), the target position of the UAV-mounted RIS is set to $(0,0,2)$. The Tx and Rx are positioned at $(-1.55,0.75,0.7)$ and $(1.65,0.77,0.69)$, respectively.
In this study, Scenario (i) serves as the baseline, as it operates in the specular reflection regime, which can result in relatively good performance even when the RIS is turned off. Scenarios (ii) and (iii) are in contrast selected for performance evaluation, as they enable a clearer assessment of the RIS contribution by ensuring that reflections from the RIS are steered toward the Rx.
All scenarios are illustrated in Fig.~\ref{fig:scenarios}.

We conduct two flights for each scenario.
In the first flight, we determine the RIS configuration in real-time by solving \eqref{eq:optProblem} for each predicted RIS pose.
This flight is divided into five phases, each with a duration of \SI{10}{\second}.
In the first, third, and fifth phase, we alternate the RIS configuration between the optimal and the off-state at a frequency of \SI{50}{\hertz} for better performance evaluation, where the off-state serves as a reference.
In addition, these phases later help to ensure accurate temporal alignment across the different datasets in post-processing.
The intermediate phases are used to continuously apply the optimal RIS configuration, i.e., without switching to the off-state in between.
In the second flight, we optimize the RIS configuration once prior flight for the respective target position and keep it constant throughout the entire flight.
Similar to the active RIS flights, the RIS configuration is additionally alternated between the optimal and off-state at the beginning and end of each flight to facilitate synchronization.

The UAV autonomously navigates to its target position in guided mode\footnote{https://ardupilot.org/copter/docs/ac2\_guidedmode.html} using the MavProxy ground station and maintains its target position once reached.
We restart the UAV and calibrate the IMUs on the flight controller before every flight to ensure the same initial conditions. The EKF state estimates are recorded on an onboard SD-card with a frequency of \SI{100}{\hertz}.
The predicted RIS position and the corresponding RIS configuration are logged on the Raspberry Pi at a frequency of \SI{50}{\hertz}.
For the hardware setup used in this work, we set the propagation interval to
$t_{\mathrm{pred}}=0.013$\si{\second}, which was determined empirically through multiple performance evaluations.

To validate the performance of the UAV-mounted RIS during the experiments, we use a vector network analyzer (VNA) of type Keysight P5026B equipped with the S9010B software option, which enables time-gating to isolate the signal component only reflected by the RIS. Specifically, a time gate spanning 61-71~\si{\nano\second} is applied, suppressing undesired multipath components and extracting only the RIS-reflected contribution at the target position. The $\mathrm{S21}$ parameter is measured over a bandwidth of \SI{650}{\mega\hertz}, spanning from \SI{5.225}{\giga\hertz} to \SI{5.875}{\giga\hertz}, with 201 frequency points, providing a frequency-resolved characterization of the channel amplitude and phase response. The measurements are conducted in \SI{50}{\hertz} intervals to ensure synchronization with the RIS switching rate as well as the update rate of the EKF-based prediction algorithm, enabling temporally aligned channel acquisition. Two VNA ports serve as the transmitter and receiver, respectively, and are each connected to a directional horn antenna of type LB-187-15-C-SF (A-Info). Within the considered frequency range, the antenna gain is at least \SI{16.35}{\deci\bel i}.

\begin{figure}[t]
	\centering
	\includegraphics[width=0.98\linewidth] {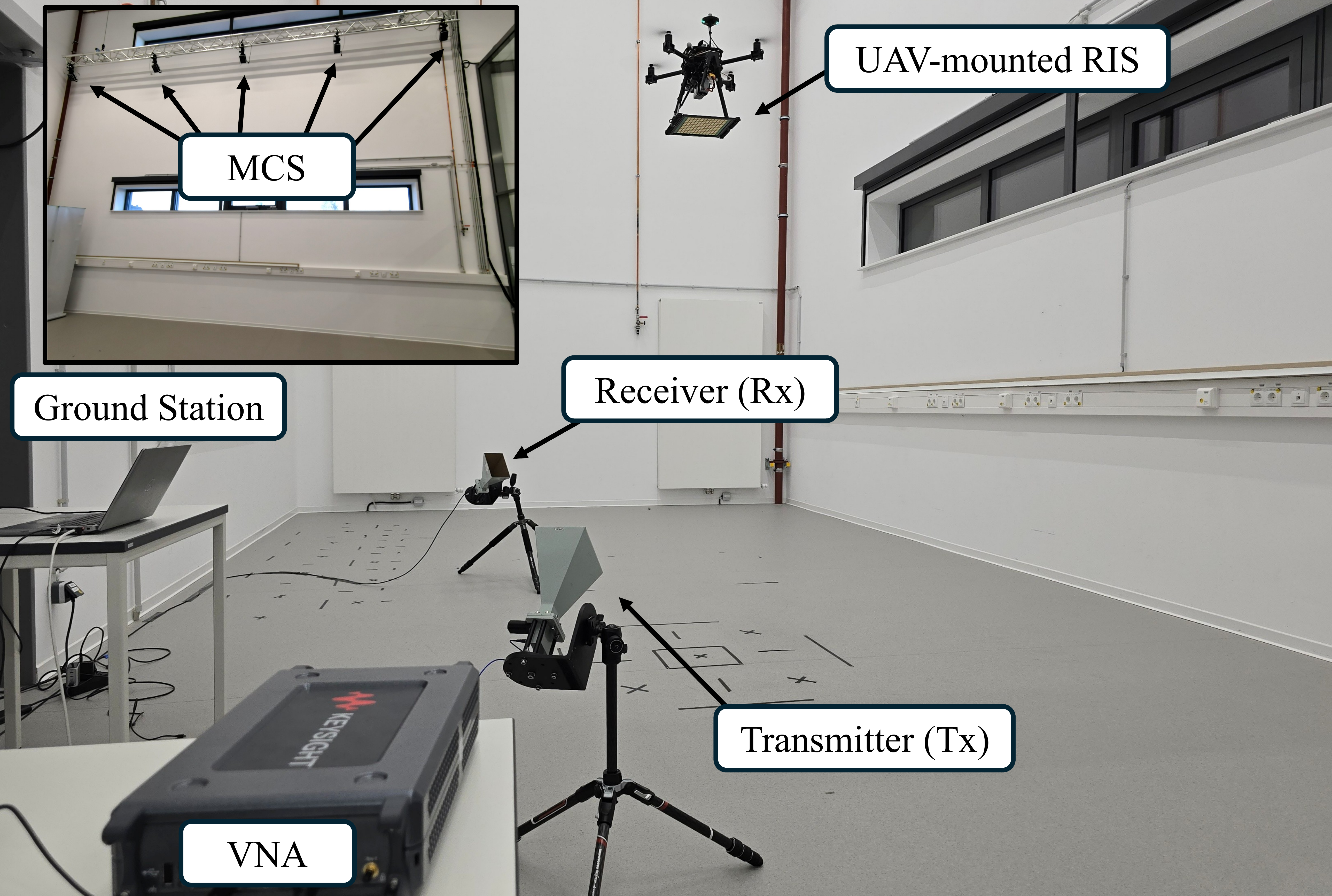}
	\caption{Experimental setup for scenario (iii). Small image in the top left corner shows the placement of the motion capture cameras.}
	\label{fig:Experiments}
\end{figure}

Reference data for the UAV position and orientation, as well as the data used for GPS substitution (see Sec.~\ref{sec:UAVSetup}), are measured and recorded using an MCS.
Specifically, 10 Vicon Vantage V5 infrared cameras and the software Tracker 3 were installed. This system is able to track objects equipped with an asymmetric pattern of reflective markers with a root mean squared error of less than \SI{0.2}{\milli\meter} with frequencies of up to 420\si{\hertz}\footnote{https://www.vicon.com/}. We calibrate the MCS once after it has reached its working temperature. The motion capture system sampling frequency is set to \SI{100}{\hertz}, i.e. the same logging frequency as the flight controller.
The complete setup for the experiment in scenario (iii) is shown in Fig.~\ref{fig:Experiments}.

We synchronize the MCS, UAV, and VNA data before evaluation to remove the temporal phase shift between the sets of data.
First, we synchronize the MCS and EKF data, after which we interpolate the MCS measurements to the Raspberry Pi timestamps.
This enables direct comparison between the predicted and the ground-truth RIS pose, which can be obtained from the MCS data using \eqref{eq:p_RIS}.
Subsequently, we discard the data recorded in the first and last 10\si{\second} of the flight, since these periods are used for takeoff and landing and are excluded from the evaluation.
Afterwards, we synchronize the estimated RIS pose with the VNA measurements.
To this end, we exploit the controlled switching between the optimized and off-states of the RIS at the beginning and end of each flight.
By logging the applied RIS configuration together with its timestamp on the Raspberry Pi, the signal quality measurements acquired by the VNA can be synchronized with the corresponding RIS configurations, ensuring accurate temporal alignment across all datasets.

\subsection{Evaluation of Pose Prediction Accuracy}
\label{sec:PredictionAccuracy}

We evaluate the accuracy of the proposed open-loop forward propagation method introduced in Sec.~\ref{sec:RISPosePred} by comparing its predictions against ground-truth data.
Furthermore, we show that directly using the most recent EKF estimates, without forward propagation, degrades RIS pose prediction performance due to the latency between state estimation and the applied RIS configuration.

\begin{figure*}[ht]
	\centering
	\includegraphics[width=1\linewidth] {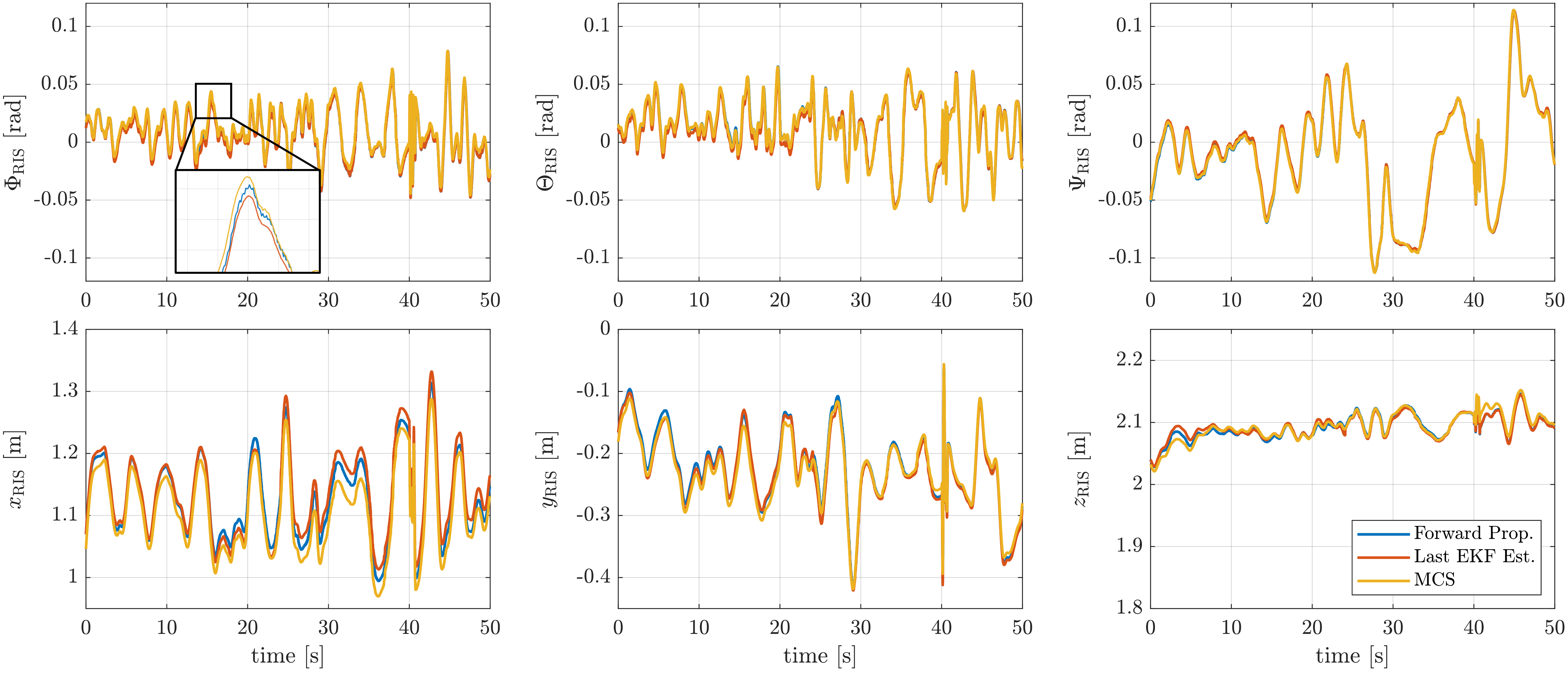}
	\caption{RIS pose prediction results for a prototype test flight. Yellow lines show the ground truth RIS pose measured by the motion capture system. Blue lines show the predicted RIS pose obtained via open-loop forward propagation, while orange lines show the RIS pose estimated using the most recent EKF estimates. First row shows the attitude, and the second row shows the position.}
	\label{fig:PredAccuracy}
\end{figure*}

For this evaluation, we conducted an additional hover flight with the UAV that was not used for the UAV-mounted RIS performance evaluation but is, in terms of positioning, identical to scenarios (i)–(iii).
The total flight duration was \SI{70}{\second}, of which \SI{50}{\second} were spent hovering.
We consider this duration to be sufficient to capture potential effects of IMU bias.
Figure~\ref{fig:PredAccuracy} shows the RIS pose measured by the MCS (yellow), the prediction obtained via open-loop forward propagation (blue), and the prediction based on the most recent EKF data (orange).
\begin{table}[b] \normalsize
    \centering
    \caption{Prediction errors for the RIS pose using open-loop forward propagation and the last EKF estimate of the UAV.}
    \begin{tabular}{l c c}
         \toprule
          &\textcolor{white}{I} Forward Propagation\textcolor{white}{I} &\textcolor{white}{I} Last EKF Est. \textcolor{white}{I}\\  
         \midrule
         $\Phi^{\mathrm{err}}_{\mathrm{avg}}$ & \SI{0.0031}{\rad} & \SI{0.0037}{\rad} \\
         $\Theta^{\mathrm{err}}_{\mathrm{avg}}$ & \SI{0.0023}{\rad} & \SI{0.0034}{\rad} \\
         $\Psi^{\mathrm{err}}_{\mathrm{avg}}$ & \SI{0.0016}{\rad} &  \SI{0.0020}{\rad} \\
         \midrule
         $\Phi^{\mathrm{err}}_{\mathrm{max}}$ & \SI{0.0094}{\rad} & \SI{0.0113}{\rad} \\
         $\Theta^{\mathrm{err}}_{\mathrm{max}}$ & \SI{0.0089}{\rad} & \SI{0.0140}{\rad} \\
         $\Psi^{\mathrm{err}}_{\mathrm{max}}$ & \SI{0.0053}{\rad} & \SI{0.0088}{\rad} \\
         \midrule
         $p^{\mathrm{err}}_{\mathrm{avg}}$ & \SI{2.35}{\centi\meter} & \SI{3.24}{\centi\meter} \\
         $p^{\mathrm{err}}_{\mathrm{max}}$ & \SI{3.88}{\centi\meter} & \SI{5.43}{\centi\meter} \\
         \bottomrule
    \end{tabular}
    \label{tab:errors}
\end{table}
As comparison metrics, we compute the average and maximum errors of the predicted roll, pitch, and yaw angles with respect to the ground truth data measured by the MCS.
More specifically, we calculate the absolute error for the roll angle $\Phi$ with
\begin{equation}\nonumber
    \Phi^{\mathrm{err}}_{k+p,\mathrm{RIS}} = |\Phi^{\mathrm{pred}}_{k+p,\mathrm{RIS}}-\Phi^{\mathrm{MCS}}_{k+p,\mathrm{RIS}}|,
\end{equation}
for all $k+p$, where $\Phi^{\mathrm{pred}}$ denotes the prediction from any of the two approaches and $\Phi^{\mathrm{MCS}}$ refers to the value from the MCS. 
The average and maximum prediction error for the roll angle can then be calculated with
\begin{equation}\nonumber
    \Phi^{\mathrm{err}}_{\mathrm{avg}} = \frac{1}{n}\sum_{k=1}^{n} \Phi^{\mathrm{err}}_{k+p,\mathrm{RIS}},
    \quad
        \Phi^{\mathrm{err}}_{\mathrm{max}} = \max_{k=1,\dots, n}(\Phi^{\mathrm{err}}_{k+p,\mathrm{RIS}}),
\end{equation}
respectively, where $n$ is the number of observations.
The errors for the pitch $\Theta$ and yaw angle $\Psi$ are defined accordingly.
Similarly, we compute the position error.
However, the spatial directions are not treated individually. Instead, we compute the Euclidean distance
\begin{equation}\nonumber
    p^{\mathrm{err}}_{k+p,\mathrm{RIS}} = \sqrt{(x^{\mathrm{err}}_{k+p,\mathrm{RIS}})^2 +(y^{\mathrm{err}}_{k+p,\mathrm{RIS}})^2 +(z^{\mathrm{err}}_{k+p,\mathrm{RIS}})^2},
\end{equation}
since for calculating the cascaded channel coefficient $h^\mathsf{casc}_m$ from \eqref{eq:chanModel}, only the absolute error is relevant.
Table~\ref{tab:errors} presents the average and maximum prediction errors for both methods.
When considering the angular rotation, it is difficult to observe noticeable differences, as the prediction errors are of similar magnitude and remain very small.
This is due to the hover flight conditions, where no large rotational angles or rapid attitude changes occur, facilitating tracking of the EKF.
In particular, using the open-loop forward propagation for predicting the RIS attitude slightly outperforms the approach based on directly using the most recent EKF data.
For $\Phi$, $\Theta$, and $\Psi$, the average estimation errors are $19\%$, $47\%$, and $25\%$ higher, respectively, when directly using the most recent EKF data.
Similar trends can be observed for the maximum estimation errors.
Although these differences appear significant in relative terms, the absolute errors between ground truth and each prediction approach remain below \SI{0.014}{\rad} across all rotational angles. In fact, they do not affect RIS optimization for the Tx–RIS/RIS–Rx distances considered in the experiments~\cite{Mueller2024}.

The errors become more pronounced when looking at the predicted position.
Predicting the RIS position using the open-loop forward propagation reduces the average prediction error by almost \SI{1}{\centi\meter}, corresponding to a relative improvement of approximately $38\%$ compared to the most recent EKF estimates. As illustrated in Fig. \ref{fig:PredAccuracy}, the most significant improvement is observed in the $x$-position.
This behavior can be attributed to the larger and more frequent position changes compared to the other spatial directions, where the inclusion of velocity and acceleration information significantly improves prediction accuracy.
Similar results can be observed for the maximum prediction error.
The open-loop forward propagation reduces the maximum position error by \SI{1.55}{\centi\meter}, corresponding to a relative improvement of nearly $40\%$.
More importantly, however, the maximum prediction error obtained by directly using the most recent EKF estimates for prediction exceeds the signal wavelength of approximately \SI{5}{\centi\meter}, which the RIS is designed for.
This implies that the resulting phase error can exceed $2\pi$, becoming too large for reliable RIS configuration and thus cannot support accurate beam steering.
Consequently, the employed optimization approach becomes invalid, and stochastic optimization methods would be required instead. 
However, the computational complexity associated with applying such methods at every time step renders real-time operation of the prototype infeasible.

The forward propagation provides sufficiently accurate predictions with respect to RIS hardware constraints, without the need for additional sensors or high-performance hardware.
Based on the observed behavior in the $x$-position, we expect the proposed approach to achieve even greater improvements over direct EKF-based prediction during trajectory flight, where the incorporation of velocity and IMU information becomes increasingly important due to rapid state changes.
However, since the EKF estimate serves as the initial state, the forward propagation may still lead to prediction errors exceeding the signal wavelength for which the RIS is designed. This could necessitate the use of additional sensors or dedicated filtering techniques for RIS pose prediction, decoupled from the UAV state estimation.

\subsection{Experimental Assessment of RIS Optimization Gains}
\label{sec:ChannelQualityInference}

In this section, we evaluate the channel quality improvements achieved through real-time optimization of the UAV-mounted RIS.
First, we compare the performance of the optimized RIS against the reference case, where the RIS is in its off-state, i.e. no elements are configured to induce phase shifts, for scenarios (i) - (iii).
Subsequently, we assess the performance gains relative to a static RIS, whose configuration was optimized offline prior to the flight for the respective target position.
\begin{figure*}[ht]
	\centering
	\includegraphics[width=0.94\linewidth] {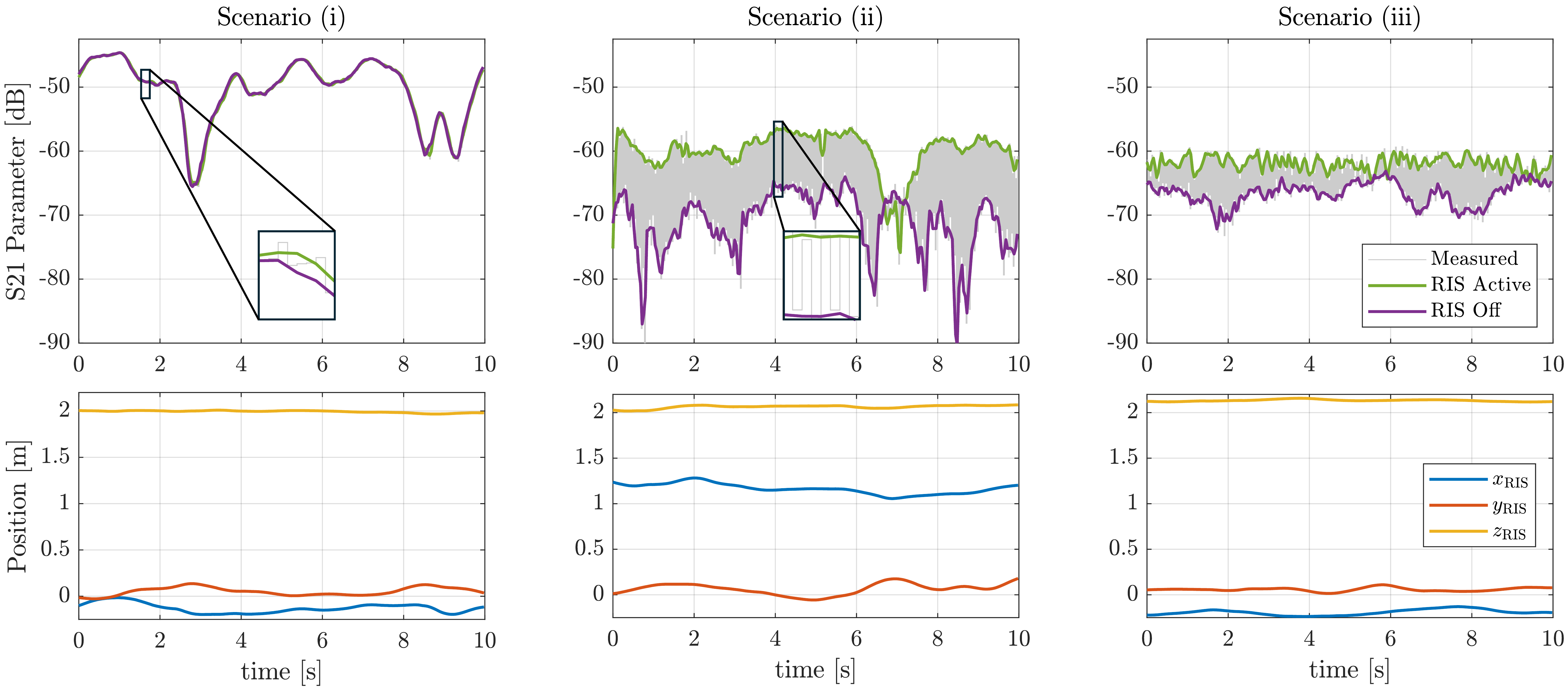}
	\caption{S21 magnitudes and RIS positions for flights corresponding to scenarios (i) - (iii). First row shows the S21 magnitudes and second row shows the RIS position for each scenario. Gray lines depict the measured S21 magnitudes, green lines the smoothed S21 magnitudes obtained with real-time RIS configuration and purple lights correspond to S21 magnitudes when the RIS is in its off-state. Blue, red, and yellow lines show the true RIS position for the $x,y,z$ position, respectively.}
	\label{fig:VNA}
\end{figure*}
Similar to the performance metrics of the prediction, we compute the average and maximum gains in the $\mathrm{S}21$ magnitude achieved by the active RIS relative to the RIS in its off-state.
Specifically, the gain in dB is defined as
\begin{equation}\nonumber
    \mathrm{S}21^{\mathrm{gain}}_{k+p} = \mathrm{S}21^{\mathrm{opt}}_{k+p}-\mathrm{S}21^{\mathrm{off}}_{k+p},
\end{equation}
for all $k+p$, where $\mathrm{S}21^{\mathrm{opt}}$ refers to the $\mathrm{S}21$ magnitude obtained through real-time RIS optimization, and $\mathrm{S}21^{\mathrm{off}}$ denotes the corresponding magnitude when the RIS is in its off-state.
The average and maximum gains can then be calculated with
\begin{equation}\label{eq:S21gains}
    \mathrm{S}21^{\mathrm{gain}}_{\mathrm{avg}} = \frac{1}{n}\sum_{k=1}^{n} \mathrm{S}21^{\mathrm{gain}}_{k+p},
    \quad
        \mathrm{S}21^{\mathrm{err}}_{\mathrm{max}} = \max_{k=1,\dots, n}(\mathrm{S}21^{\mathrm{gain}}_{k+p}),
\end{equation}
where $n$ is the number of observations.
In addition, we compute the average $\mathrm{S}21$ magnitude for each scenario, independent of the off-state reference, as
\begin{equation}\nonumber
    \mathrm{S}21_{\mathrm{avg}} = \frac{1}{n}\sum_{k=1}^{n} \mathrm{S}21_{k+p}
\end{equation}
to quantify the impact of different deployment locations on the performance of the UAV-mounted RIS.
For the first part of the evaluation, we consider the flight segments during which the RIS configuration is alternated.
\begin{table}[b] \normalsize
    \centering
    \caption{S21 magnitude and gains for scenarios (i)–(iii).}
    \begin{tabular}{l c c c}
         \toprule
          Scenario &\textcolor{white}{MM} (i) \textcolor{white}{MM} &\textcolor{white}{MM} (ii) \textcolor{white}{MM} & \textcolor{white}{MM} (iii) \textcolor{white}{MM} \\  
         \midrule
         $\mathrm{S}21_{\mathrm{avg}}$ & \SI{-50.50}{\deci\bel} & \SI{-60.05}{\deci\bel} & \SI{-61.92}{\deci\bel} \\
         $\mathrm{S}21^{\mathrm{gain}}_{\mathrm{avg}}$ & \SI{0.05}{\deci\bel} & \SI{11.24}{\deci\bel} & \SI{4.73}{\deci\bel} \\
         $\mathrm{S}21^{\mathrm{gain}}_{\mathrm{max}}$ & \SI{1.50}{\deci\bel} & \SI{36.26}{\deci\bel} & \SI{12.65}{\deci\bel} \\
         \bottomrule
    \end{tabular}
    \label{tab:VNA}
\end{table}

Figure~\ref{fig:VNA} shows the $\mathrm{S}21$ magnitudes during representative periods of RIS configuration alternation in each scenario.
In addition, the UAV position at the corresponding time step is shown.
We omit plotting the RIS attitude, as it has a negligible impact on the channel quality for the attitude changes encountered during hover flight.
Table~\ref{tab:VNA} presents the average and maximum magnitude gains as computed with \eqref{eq:S21gains}.
Furthermore, it shows the average $\mathrm{S}21$ magnitude obtained when continuously applying the optimal RIS configuration.

When comparing the average channel gains $\mathrm{S}21^{\mathrm{gain}}_{\mathrm{avg}}$ in Tab.~\ref{tab:VNA}, we can observe that the active RIS consistently outperforms the RIS in its off-state, signified by the positive gain values.
As expected, the performance difference is more pronounced in scenario (ii) and (iii), when compared to scenario (i).
As discussed in Section~\ref{sec:Experiments}, scenario~(i) effectively reduces to the specular reflection case for the considered system geometry, where the optimal RIS configuration coincides with the RIS off-state due to the lower imposed attenuation.
Consequently, only minor differences can be observed, where the RIS effectively counteracts disturbances, that cause deviations of the UAV from its target position.
For scenarios (ii) and (iii), however, we can observe significant performance differences between the active and the turned-off RIS, thereby experimentally validating both the effectiveness of the proposed optimization framework and the functionality of the overall experimental setup.
More precisely, the active RIS achieves an average gain of \SI{11.24}{\deci\bel} over the turned-off RIS in scenario (ii), while the corresponding gain in scenario (iii) is \SI{4.73}{\deci\bel}.
Similar results are obtained for the maximum achievable gain relative to the off configuration, with differences of up to \SI{36.26}{\deci\bel} in scenario (ii) and \SI{12.65}{\deci\bel} in scenario (iii).
These maximum gains occur when the UAV significantly deviates from its target position, during which small attitude variations are also induced as the UAV attempts to maintain its target position.
In such situations, actively reconfiguring the RIS becomes crucial to counteract disturbances, as passive specular reflections alone are no longer sufficient to maintain a reliable link to the receiver.

\begin{figure}[t]
	\centering
	\includegraphics[width=0.98\linewidth] {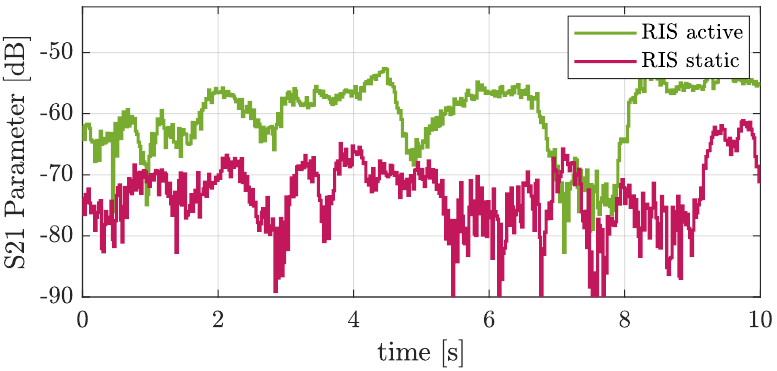}
	\caption{S21 magnitudes for a flight in scenario (ii) with real-time RIS configuration (green line) and a static configuration (pink line) optimized for the respective target position.}
	\label{fig:staticactive}
\end{figure}

In scenario (i), the optimal RIS configuration coincides with the RIS off-state (see Fig.~\ref{fig:VNA}), which is identical to the static configuration optimized for the target position.
This raises the question of whether real-time adaptation of the RIS is necessary, as a static configuration may already be sufficient.
To further investigate this, we compare two flights corresponding to scenario (ii), for which the UAV movement patterns are nearly identical.
The results are shown in Fig.~\ref{fig:staticactive}.
It becomes evident that, in the presence of uncertainties and disturbances, a static RIS configuration is insufficient. More specifically, the active UAV-mounted RIS achieves an $\mathrm{S}21$ magnitude of \SI{-60.25}{\deci\bel}, whereas the static RIS configuration only reaches \SI{-73.41}{\deci\bel}, which is on the same order as the RIS in its off-state when compared to the results in Fig.~\ref{fig:VNA}.
From a more general perspective, the proposed active optimization framework constitutes a general RIS configuration solution, which in the specular reflection regime reduces to the off-state as a special case. Accordingly, although a static RIS configuration may achieve comparable performance in specific scenarios such as scenario (i), we have shown that real-time reconfiguration is generally required to ensure sufficient channel quality in practical scenarios.

While real-time configuration of the UAV-mounted RIS, provides significant performance gains, its deployment position plays an equally crucial role in attaining a good channel link quality.
Positioning the RIS at the midpoint between the transmitter and receiver in the $x$–$y$-plane yields the highest magnitudes, as multipath fading is reduced when the Tx–RIS–Rx path length is minimized (see \eqref{eq:chanModel}).
More specifically, the $\mathrm{S}21$ magnitude is approximately \SI{10}{\deci\bel} higher in the mid-position compared to the other deployment locations.
However, in real-world missions, the RIS cannot always be positioned optimally, and thus alternative deployment locations must also be considered.
When comparing scenarios (ii) and (iii), the average $\mathrm{S}21$ magnitude over the observed interval is found to be nearly identical.
This is notable, as in scenario (ii) the RIS is still located along the direct line between the transmitter and receiver, whereas in scenario (iii) the RIS is laterally displaced in $x$ and $y$.
With respect to the same $z$-position, this results in increased Tx–RIS–Rx path lengths and a significantly degraded angle-of-arrival/-departure geometry.
These observations further highlight the performance potential of a real-time reconfigurable UAV-mounted RIS, as it is able to maintain substantial channel gains even under unfavorable deployment conditions.

\section{Conclusion}

In this work, we presented the first fully functional real-time reconfigurable UAV-mounted RIS prototype and validated its performance under real-world conditions.
We demonstrated the feasibility of real-time UAV-mounted RIS and quantified the resulting improvements in the channel link quality, closing the existing research gap between predominantly theoretical studies and practical implementation.
Experimental evaluation across multiple deployment scenarios showed that the proposed system consistently outperforms a static RIS configuration while effectively compensating for disturbances in dynamic environments.
The results further indicate that positioning the RIS within the specular reflection regime between transmitter and receiver maximizes the $\mathrm{S}_{21}$ magnitude for both static and active RIS configurations.
Nevertheless, the active RIS exhibits only minor performance degradation when deployed away from this optimal position, highlighting its robustness against non-ideal placement.
Future work will investigate UAV-mounted RIS during trajectory flight, where continuous motion introduces reduced state estimation accuracy and rapidly varying wireless channels, posing additional challenges for channel prediction and real-time RIS control.

\bibliographystyle{IEEEtran}
\bibliography{bib}

\vspace{-1.0cm}

\begin{IEEEbiography}[{\includegraphics[width=1in,height=1.25in,clip,keepaspectratio]{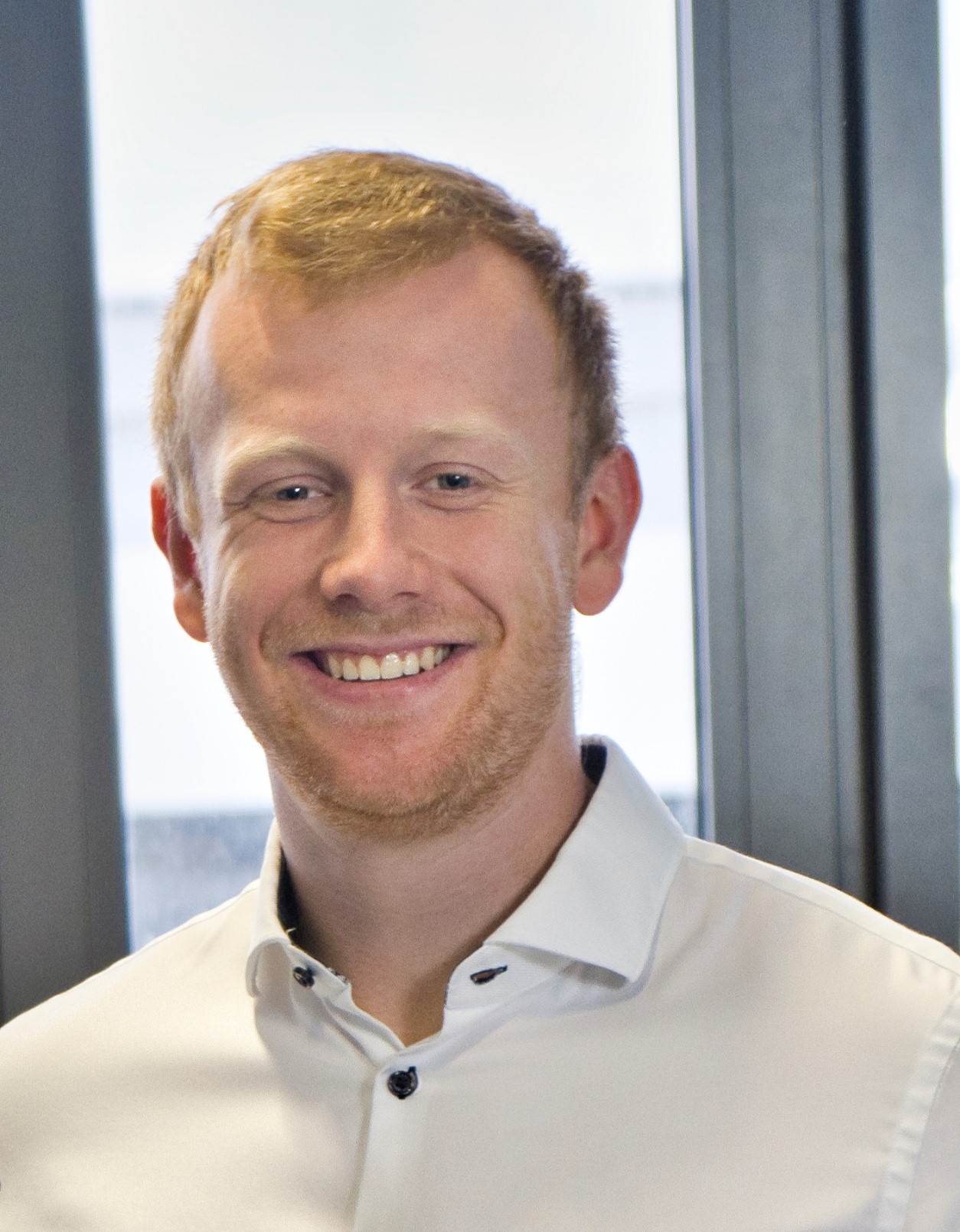}}]{David Mueller} received the B.Sc. and M.Sc. in mechanical engineering from Ruhr-Universität Bochum, Germany, in 2021 and 2023, respectively, where he is currently pursuing the Ph.D. degree with the chair of Automatic Control and Systems Theory. His research interests include state estimation, system integration, and uncertainties in measurement with focus on UAV applications.
\end{IEEEbiography}

\vspace{-1.0cm}

\vspace{11pt}

\begin{IEEEbiography}[{\includegraphics[width=1in,height=1.25in,clip,keepaspectratio]{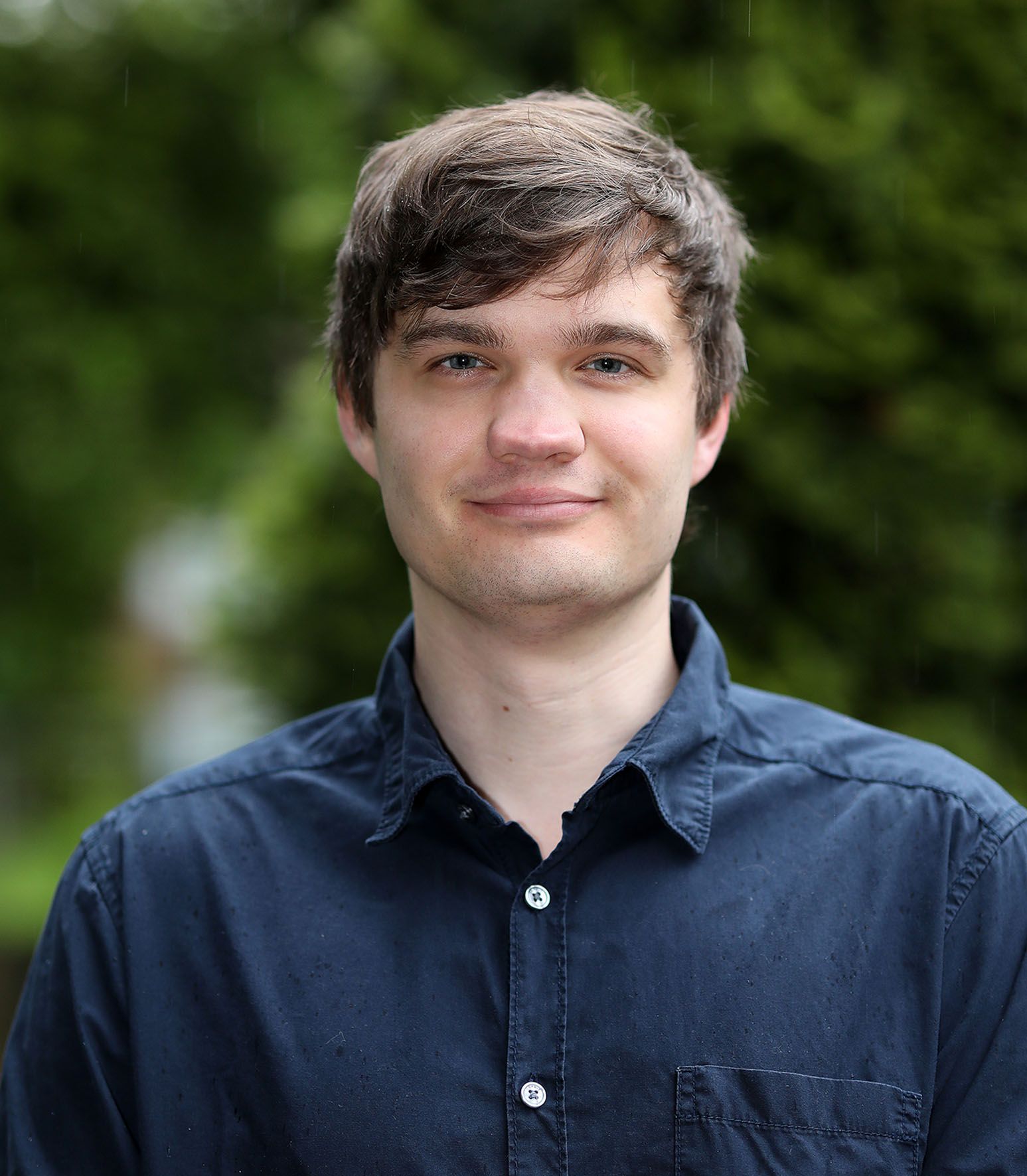}}]{Kevin Weinberger} received the B.Sc. and M.Sc. degrees in electrical engineering and information technology from Ruhr-Universität Bochum, Germany, in 2017 and 2020, respectively, where he is currently pursuing the Ph.D. degree with the Institute of Digital Communication Systems. His research focuses on wireless communications, with particular emphasis on information-theoretic methods and signal processing for next-generation communication systems. His work contributes to the analysis and design of advanced wireless networks, including performance characterization and system-level modeling of communication links under realistic propagation conditions.
\end{IEEEbiography}

\vspace{-1.0cm}

\vspace{11pt}

\begin{IEEEbiography}[{\includegraphics[width=1in,height=1.25in,clip,keepaspectratio]{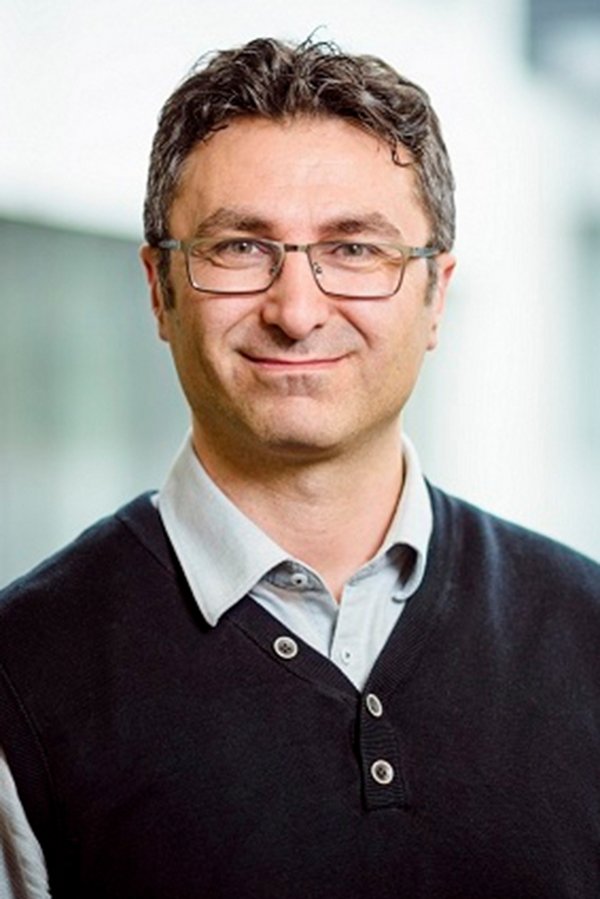}}]{Aydin Sezgin} (Senior Member, IEEE) received
the Dr. Ing. (Ph.D.) degree in electrical engineering from TU Berlin, in 2005. From 2001 to 2006,
he was with the Heinrich-Hertz-Institut, Berlin.
From 2006 to 2008, he held a postdoctoral position, and was also a Lecturer with the Information
Systems Laboratory, Department of Electrical
Engineering, Stanford University, Stanford, CA,
USA. From 2008 to 2009, he held a postdoctoral
position with the Department of Electrical Engineering and Computer Science, University of California, Irvine, CA, USA.
From 2009 to 2011, he was the Head of the Emmy-Noether-Research Group
on Wireless Networks, Ulm University. In 2011, he joined TU Darmstadt,
Germany, as a Professor. He is currently a Professor with Ruhr University
Bochum, Germany. He has published several book chapters, more than
70 journals and 200 conference papers in these topics. 
He is a winner of the ITG-Sponsorship Award, in 2006. He was a First Recipient of the prestigious Emmy-Noether Grant by the German Research Foundation in communication engineering, in 2009.
\end{IEEEbiography}

\vspace{-1.0cm}

\vspace{11pt}

\begin{IEEEbiography}[{\includegraphics[width=1in,height=1.25in,clip,keepaspectratio]{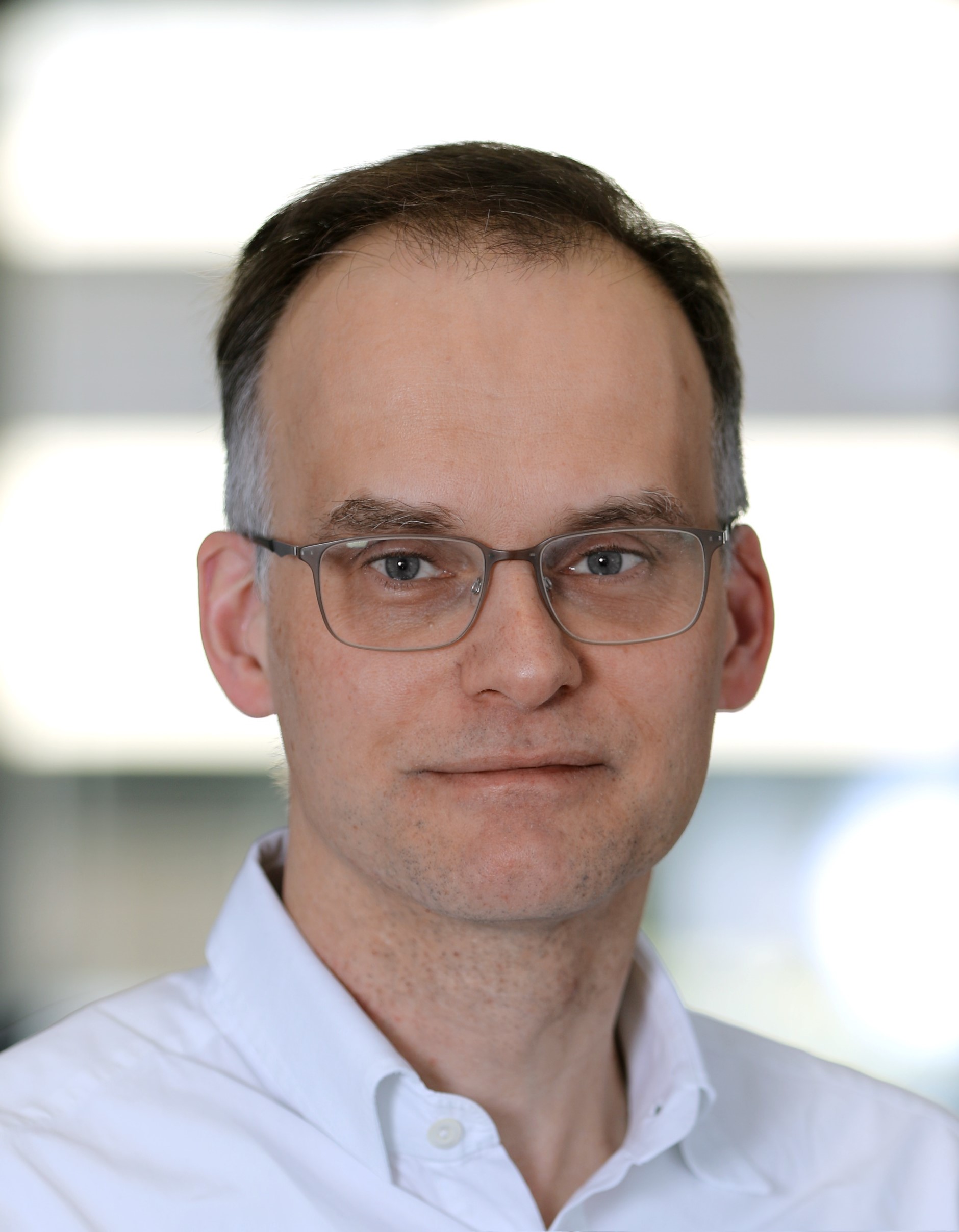}}]{Martin Moennigmann} received the Diploma and Ph.D. degrees from RWTH Aachen University, Germany, in 1998 and 2004, respectively. He was a postdoctoral research scholar at Princeton University 2004–2005, deputy director of the Aachen Institute for Advanced Study in Computational Engineering 2005–2007 and assistant professor at Technische Universität Braunschweig 2007–2009, Germany. In 2009 he was appointed full professor and head of the chair for Automatic Control and Systems Theory at Ruhr-Universität Bochum, Germany. His research interests include predictive and optimal control, nonlinear dynamics, and applications in the energy and chemical engineering sectors
\end{IEEEbiography}

\vfill

\end{document}